# Enhancing resilience in primary care through inter-regional patient flow network

Lu Zhong[1,2,3,*], Lior Rennert[1], Sen Pei[4], Jianxi Gao[2,3,*]

[1]Department of Public Health Sciences, Clemson University, Clemson, SC, USA, 29631

[2]Department of Computer Science, Rensselaer Polytechnic Institute, Troy, NY, USA 12180

[3]Network Science and Technology Center, Rensselaer Polytechnic Institute, Troy, NY, USA 12180

[4]Department of Environmental Health Sciences, Columbia University, New York, NY, USA 10032

*Corresponding Author: Lu Zhong, Jianxi Gao

**Email**: lucinezhong@gmail.com, gaoj8@rpi.edu

**Abstract**

Large-scale disasters, such as pandemics and climate-related events, place extraordinary pressure on healthcare providers due to extreme demand surges. Managing these surges is essential to sustaining healthcare resilience. Although numerous studies on healthcare resilience, far less attention has been given to physicians and to how patterns of patient movement can help redistribute demand and alleviate stress on overburdened providers. In this study, we analyzed billions of electronic medical records documenting patient visits to primary care physicians (PCPs) to construct inter-regional patient flow networks across the U.S. During the COVID-19 pandemic, we observed that cross-regional flow rose to 2.81%, compared to the pre-pandemic level of 2.08%. This redistribution absorbed, on average, 58% of the excess stress on PCPs, meaning more than half of the surging demand was handled by patients' moves to less burdened regions, an absolute 43 percentage point improvement from the pre-pandemic baseline of 15%. Further analysis suggests that strengthening cross-regional patient flow could allow the healthcare system to absorb even more stress and reduce the demand for PCPs. These findings provide structural insights for the healthcare system to enhance its pandemic preparedness and disaster responses, and to improve patient care during crises.

## Introduction

Large-scale disasters, such as extreme weather events driven by climate change and global pandemics, continually challenge the resilience of healthcare systems[1–3], that is, the ability of system to ensuring patient care under extreme demand surges. At the center of this resilience are physicians[4–6], who play a pivotal role in sustaining care delivery under crisis conditions. The COVID-19 pandemic, for example, placed unprecedented pressure on the clinical workforce [7–9]. As patient demand rapidly exceeded available capacity, physicians were required to manage intensified workloads, extended working hours, resource scarcity, and complex triage decisions. Their capacity to adapt, prioritize care, and coordinate limited resources became critical to maintaining system functionality. However, when demand overwhelms local physician capacity, individual adaptability alone is insufficient to main resilience. Recognizing this constrain, some jurisdictions, such as Arizona and New York in the United States[10,11], London in the United Kingdom[9,12], and various other countries[13,14], implemented proactive, system-level strategies. These included the establishment of centralized surge coordination systems (e.g., "surge lines"), cross-regional collaborations across, and structured patient transfers across healthcare networks [13,15,16]. The underlying goal was to utilize neighbor resources, that is, to redistribute patients from regions where local physicians or healthcare resources were overburdened to regions where physicians had available capacity. Such structural responses have been widely acknowledged as effective in mitigating provider overload, improving access to timely care, enhancing systemic responsiveness, and ultimately saving lives[15,17,18].

Despite their intuitive appeal, the extent to which cross-regional patient movement improve healthcare resilience remains insufficiently understood. Much existing research on healthcare resilience has primarily emphasized system-wide indicators[4,19–21], such as patient visits[4], staffing levels[22,23], or quality-of-care [24,25], to quantify how systems absorb stress and recovery back to normal. While informative, these approaches largely overlook how stress is distributed across individual physicians and how geographic patient redistribution may help rebalance clinical workloads. Emerging evidence suggests that patient movements can influence outcome [26,27] ; for instance, studies have shown that cross-facilities movements can improve clinical outcomes for patients with brain metastases and their mortality. Whether such movements systematically drive resilience, however, remains poorly characterized.

A structural, network-based perspective [28,29] offers a way forward. Rather than focusing solely on aggregate indicators, this approach emphasizes on sustaining patient care through neighboring providers[14,30–32]. Although theories of network resilience are well developed in fields such as engineering, business, and ecology[33–35], their translation into healthcare has been limited. As a result, resilience in healthcare has remained largely conceptual, with few actionable strategies to guide system design[3,36]. The growing availability of electronic health records (EHRs) now makes it possible to model healthcare systems as provider-based networks and empirically examine how patient movements shapes continuity of care and outcomes[37].

In this study, we construct cross-regional patient visits to PCPs across U.S. states at the three-digit ZIP code level using millions of electronic medical records (2017–2022). In our framework, incoming patient volume represents demand, and the number of PCPs at each region represent treatment capacity within each region. We quantify excess stress when demand exceeds the capacity of physicians practicing in a region. To measure the system's ability to mitigate this stress, we introduce absorptivity—the proportion of excess stress alleviated through patient movement to physicians in other regions. We find that absorptivity rose significantly during the pandemic, though the extent varied by region size. Further analysis suggests that regions that more strongly connected through patient flows have higher absorptivity, lower clinician demand, and better expand healthcare systems' capacity to withstand future shocks. By grounding healthcare resilience observed pattern of patient movement across regions, our work moves beyond system-wide indicators to reveal actionable structural levers. Our findings highlight the critical role of network-aware strategies in safeguarding access to care and improving outcomes during crises, providing a foundation for resilience-informed system design and cross-regional response in future public health emergencies.

## Results

**Structural shifts in patient flow networks**. Patients often cross regional boundaries to access healthcare, driven by medical needs, service availability, hospital referrals, and adherence to crisis regulations and policies. Using electronic medical record (EMR) data capturing patient visits to primary care physicians (PCPs) in sub-regions (defined by three-digit ZIP codes), we constructed cross-regional networks of patient visit flows (Fig. 1a–b). In these networks, each node represents a sub-region, and directed edges represent patient movements between sub-regions. Edge weights correspond to the number of patients who shifted their visits from a PCP in one sub-region to a PCP in another (see Methods). We compared these networks over two phases: pre-pandemic (2017-2019) and during-pandemic (2020-2022). As demonstrated in Table 1, across the 37 states analyzed, the proportion of cross-region flows ($\sigma$) increased from 2.08% [95% CI, 1.62%-2.54 %] to 2.81% [95% CI, 2.34%-3.29 %] during the pandemic, an increase of 0.74% [P-value=0.0021]. To further examine heterogeneity, we classified sub-regions by population size into small (<50,000), medium (50,000-100,000), large (100,000-1,000,000), and mega-sized categories (>1,000,000 population). Before the pandemic, cross-regional patient flows were more common in small and medium regions, suggesting greater reliance on larger or urban regions for healthcare services. During the pandemic, however, cross-regional flows from small and medium regions declined, while flows involving large and mega-sized regions increased. This shift suggests that the pandemic restructured patient movement patterns, concentrating demand toward larger and more urban regions where service availability and provider capacity were relatively greater despite widespread disruptions.

Beyond volume, the structures of the patient flow networks shifted markedly. Figure 1a-b illustrates the spatial distribution of flow networks across the two distinct periods. During the pandemic, networks became denser, with an average of 0.08 [P-value = 0.0028] more connections per region compared to pre-pandemic levels. For example, the zip-cod region in New York, the out-degree rose from 4 to 18. When stratified by region size, mega-sized regions exhibited the most pronounced increases in out-degree (out-connection) with 0.61 [P-value=0.0014], indicating that larger urban hubs became more centrally connected during the pandemic. Geographic reach also expanded. The average travel distance for cross-regional patient visits increased from 38.34 km pre-pandemic to 42.07 km during the pandemic, an average increase of 3.73 km [P-value=0.0001]. This increase was particularly notable for larger regions: large regions experienced an average increase of 3.75 km, while mega-sized regions saw an even greater increase of 9.69 km. Together, these findings indicate that the pandemic catalyzed a large-scale reconfiguration of patient movements, characterized by expansion of geographic reach and reinforcement of central hubs.

**Absorptivity: quantifying excess stress alleviation.** To assess whether these structural shifts improved resilience, we developed a network-based metric termed absorptivity (r), defined as the proportion of excess healthcare stress that can be alleviated through cross-regional patient redistribution. As illustrated in Fig. 2, for a focal region *i* (node A), when stress exceeds its local capacity, the excessive stress $\lambda_i(t)$ is redistributed to neighboring regions $j$ (nodes B, C, and D) through patient flows, with resultant stress $\lambda_i^W(t)$. Absorptivity is defined as the magnitude of stress reduction achieved through these patient flows $r(t) = 1 - \frac{\lambda_i^W(t)}{\lambda_i(t)}$, which captures the proportion of excess stress alleviated. A higher absorptivity value of $r(t)$ indicates that a larger share of stress has been successfully absorbed.

We here define healthcare stress as the volume of incoming patients recorded in the EMR dataset (see Methods) and measure regional treatment capacity as the number of patients that local PCPs can manage. Then the excess stress is the stress greater than treatment capacity. Figures S1–S2 illustrate the overall ratio of patient over physician and the spatial and temporal distribution of excess stress during pandemics. Although overall healthcare utilization declined during the pandemic, the shortage of available physicians also contributed to excess stress in certain regions. The pronounced spatial heterogeneity further motivates us to quantify whether patient movements toward less-burdened regions occurred, and to assess how much absorptive capacity was gained through cross-regional movements.

**Temporal pattern of excess stress and absorptivity.** We construct a counterfactual scenario by fixing the pre-pandemic network structure as the baseline configuration. Specifically, we use the pre-pandemic patient flow network to simulate how demand would have been redistributed under pandemic-level stress, holding network topology constant. Outcomes observed under the during-pandemic network are then compared against this baseline to quantify the extent to which structural changes in patient flow contributed to differences in stress distribution and absorptive capacity.

Figure 3 shows the seasonally excess stress normalized their capacity, stratified by sub-region size. Small and medium regions experience higher and more volatile stress relative to treatment capacity, with pronounced spikes during peak pandemic waves (Fig 3a-b). In contrast, large regions show more moderate and gradually increasing stress (Fig 3c), while mega regions consistently maintain the lowest stress levels (Fig 3d). These findings indicate that healthcare demand–capacity imbalances were unevenly distributed, with smaller regions more vulnerable to sudden surges in patient load. The lower panel of each subfigure shows the absorptivity over time. During-pandemic networks consistently demonstrate higher absorptivity compared with pre-pandemic networks across all regions, indicating improved capacity to redistribute excess demand. The increase is particularly pronounced in mega and medium regions. Smaller regions benefit from these network adaptations but exhibit more variability in absorptive capacity. Overall, the results suggest that structural changes in patient flow networks during the pandemic enhanced the system's ability to buffer localized stress through inter-regional redistribution.

**Spatial distribution of excessive stress and absorptivity.** We aggregated the outcomes over the 2020–2022 period to compare the pre-pandemic and during-pandemic networks. Overall, the pre-pandemic network exhibited an absorptivity of 0.15, whereas the during-pandemic network showed a substantially higher absorptivity of 0.58, an increase of 0.43 (P-value = 0.0001). Stratified by region size, absorptivity during the pandemic was lowest in small regions (0.56), followed by large regions (0.57), medium regions (0.60), and highest in mega-sized regions (0.66). The magnitude of improvement relative to the pre-pandemic baseline was greatest in mega regions. These results suggest that although overall absorptivity improved during the pandemic, the extent of improvement varied across regions.

Figure 4a shows the spatial distribution of stress across all regions. While unnormalized excess stress followed a log-normal distribution (Fig. 3b), normalization revealed a power-law distribution (Fig. 3c), indicating uneven stress burdens. Analysis of absorptivity revealed a bimodal distribution (Fig. 3d) and a clear shift toward higher values during the pandemic. In the pre-pandemic network, most regions had absorptivity values near zero: 64.3% fell below $r$< 0.1, and only 3.4% exceeded $r$ > 0.9. By contrast, during the pandemic just 32.9% fell below $r$ < 0.1, while 52.3% exceeded $r$> 0.9, reflecting improved network performance and greater absorptivity under stress. By region size (Fig. 3e–g), mega-regions have the largest absolute volumes of patient demand, but medium-sized and small-sized regions bore the greatest normalized burden relative to treatment capacity. For the value of absorptive, the proportion of regions with $r$ < 0.1 was lowest in small-sized regions (25.3%) but higher in medium-sized (29.5%), large (34.6%), and mega-regions (26.7%). The share of regions achieving $r$ > 0.9 was lowest in small-sized regions (42.2%) but higher in medium-sized (53.8%), large (51.2%), and mega-regions (58.6%). This bimodality suggests polarization, that is, regions tended to cluster into either low-absorptive or highly absorptive groups, reflecting unequal structural connectivity and surrounding capacity.

**Implications of strengthened cross-regional patient movement.** To estimate maximal achievable stress redistribution, we simulated expanded patient flows using a broadcasting approach, allowing flows from high-stress regions to nearby areas with available capacity.

Figures 5a–b illustrate two scenarios in which patient flows were constrained to within 100 km and 150 km. Compared with observed during-pandemic networks, these simulated networks expanded substantially, adding 523 new cross-regional connections under the 100 km constraint (a 56% increase) and 2,060 new connections under the 150 km constraint (a 220% increase). These additional connections translated into significant resilience gains: absorptivity increased to 0.73 (a 25.9% rise) under the 100 km scenario and to 0.88 (a 50% rise) under the 150 km scenario (Fig. 5c). The improvements were widespread but uneven. Under the 100 km constraint, 8 states achieved absorptivity levels above $r$ > 0.9, while in the 150 km scenario that number nearly doubled to 21 states. However, geographically isolated states such as Wyoming showed no improvement, reflecting the limits of patient flow networks in regions with sparse connectivity and limited neighboring capacity.

However, regional collaboration alone is insufficient in fully alleviate stress ($r = 1$). To achieve full absorptive capacity, additional physicians are required to expand local treatment resources. Figure 5d illustrates the additional average number of physicians required across seasons to achieve full absorptivity ($r = 1$) in each state. Compared with during-pandemic network, far fewer physicians were required in the simulated networks, that is, 1,600 (38.7% fewer) physicians needed under the 100 km scenario and 907 physicians (65.2% fewer) under the 150 km scenario, compared with 2,610 physicians in the during-pandemic networks. These results suggest that strengthening patient flow networks not only enhances absorptivity but also reduces additional need on physician supply.

**Discussion**

Our study highlights the transformative role of patient movements in mitigating healthcare stress during the COVID-19 pandemic. By using electronic medical records that document patient visits to PCPs at the three-digit ZIP code regions, we mapped inter-regional patient flow network and identified significant structural shifts. During the pandemic, cross-regional flows increased to 2.81% (0.74 percentage points above pre-pandemic levels), with mega-regions emerging as key hubs for redistributing healthcare stress. These structural shifts translated into measurable resilience benefits, i.e., absorptivity, a network-based metric quantifying the proportion of excess system stress alleviated through redistribution of patients. On average, absorptivity rose to 0.58 during the pandemic, indicating that networks absorbed more than half of excess stress, a 43% improvement compared with the pre-pandemic baseline Simulation experiments further suggest that strengthening regional coordination could increase absorptivity to 0.73–0.88 while substantially reducing additional physician demand.

However, absorptive capacity was not evenly distributed. Medium- and large-sized regions exhibited a higher prevalence of high-absorptivity areas compared with mega regions, reflecting structural inequalities in connectivity and surrounding capacity. Small regions, although often vulnerable to capacity constraints, sometimes achieved moderate absorptivity when effectively linked to nearby hubs. These findings underscore that resilience is not solely a function of local capacity but also of spatial position and inter-regional connectivity.

While our simulations modeled the aspirational condition of full absorptivity ($r = 1$), several practical constraints make this difficult to achieve. Geographic isolation and sparse connectivity restrict redistribution potential in rural or frontier regions. Persistent workforce shortages continue to act as a fundamental bottleneck, since even well-designed networks cannot operate without adequate staff. Infrastructure and data interoperability also serve as critical prerequisites; without real-time information exchange and coordinated transfers, networks may fail to redistribute stress effectively. Nonetheless, our study demonstrates that even short of this ideal, patient flow networks can stabilize access to care and ease workforce pressures during crises.

For primary care, these findings highlight that resilience extends beyond individual practice capacity. PCP networks function as distributed systems, where patient-sharing relationships enable stress buffering during crises[38]. Strengthening referral pathways, regional coordination agreements, and use the telemedicine could substantially enhance absorptive capacity without proportionally increasing workforce supply[23,38–41]. Several limitations should be acknowledged. First, our analysis relies on EMR data, which may be subject to underreporting or inconsistent documentation, potentially biasing estimates of stress and capacity[42]. Second, our analysis does not explicitly incorporate operational forms of institutional collaboration, such as referral patterns, ICU bed-sharing[43], equipment logistics, or patient preferences, factors that are essential for fully capturing network functionality[19,44]. Third, due to dataset limitations, our analysis is restricted to three-digit ZIP code regions. This coarse geographic aggregation reduces spatial resolution and may obscure meaningful heterogeneity in patient movement patterns and local healthcare capacity. Despite these limitations, our study advances structural understanding of healthcare resilience for PCP networks under stress. As global health systems face intensifying threats from pandemics, climate-driven disasters, and geopolitical instability, enhancing cross-regional or cross-facility coordination will be central to ensuring resilient and equitable care delivery in the future.

## Methods

**Dataset.** The dataset used in this study is the Healthjump dataset, sourced from the COVID-19 Research Database[45]. Healthjump is a data integration platform that consolidates electronic medical records and practice management systems. The Healthjump network aggregates data from over 70,000 hospitals and clinics and more than 1,500 healthcare organizations across all U.S. states. The dataset represents encounter data captured at the point of care and is not derived from payors or insurance claims. Although the data are sourced from different institutions, the dataset only includes patient-provider encounter records rather than institutional-level operational data. Specifically, it contains detailed information on patient diagnoses, procedures, encounters, medical histories, residential areas (three-digit zip codes), and provider characteristics, including whether the provider is a primary care physician (PCP). We here extract the dataset for visits for PCPs and remove incomplete records. Table S1 presents an overview of the records on primary care from 2017 to 2022.

**Ethics statement.** Ethical approval was not required for this study, as it relied exclusively on fully anonymized data sourced from the Healthjump EMR database. The dataset complies with the privacy standards set by the Health Insurance Portability and Accountability Act (HIPAA) of 1996, with safeguards in place through agreements with both healthcare providers and patients. Because the data contains no personally identifiable information, the study qualifies for exemption from institutional ethical review.

**Construction of the Patient Flow Network**. According to literature[27], we define patient movement between regions using sequential visits to PCPs. If a patient $k$ initially visits a PCP at region $i$ at time $t$ and subsequently visits a PCP at region $j$ at time $t+v$, this sequence is considered a transit between these regions. We define the indicator variable as $w_{ij}^{k}(t) = 1$ otherwise $w_{ij}^{k}(t) = 0$. Here, $v$ represents the time window (we set $v \leq 3$ months). The total patient flow weight $w_{ij}(t)$ from region $i$ to region $j$ is defined as the number of patients moving between two regions:

$$w_{ij}(t) = \sum_k w_{ij}^{k}(t) \tag{1}$$

The ratio of cross-region patient flow is defined as

$$\sigma(t) = \frac{\emptyset(t)}{\theta(t)} = \frac{\sum_{i\neq j} w_{ij}(t)}{\sum_{i,j} w_{ij}(t)} \tag{2}$$

A higher value $\sigma(t)$ indicates that more patients move across regions.

In addition, for the cross-regional flow network, we calculate the out-degree of each region at time $t$, defined as the number of outgoing connections to other regions. We also measure the spatial travel distance associated with these outgoing connections to capture the geographic extent of patient mobility.

As the dataset expands over time due to continuous data collection, adjustments are necessary to ensure comparability across periods. To address this, we apply a threshold to filter cross-regional flows. For the pre-pandemic period, we use the three-year aggregated average (2017–2019) to represent baseline flow weights, retaining only edges with weights greater than 1 as valid connection. Let $\theta(Pre)$ denote the total number of patient visits during the pre-pandemic period. During the pandemic period at time $t$, only edges with weights greater than $\theta(t)/\theta(Pre)$ are considered valid. This normalization ensures that differences in total visit volume over time do not bias the comparison of network structure.

**Measuring excess healthcare stress at regions.** For region $i$ at time $t$, let $\rho_i(t)$ denotes the expected patient-to-physician ratio, which serves as a benchmark for the region's treatment capacity under normal operating conditions. Given the number of physicians $P_i(t)$, the treatment capacity is defined as $C_i(t) = P_i(t)\,\rho_i(t)$, representing the maximum number of patients that can be treated. Let $S_i(t)$ represent the number of incoming patients to region $i$ at time $t$. Excessive stress is then defined as

$$\lambda_i(t) = \max\left(S_i(t) - C_i(t), 0\right) \quad (3)$$

If $\lambda_i(t) > 0$, the region is over-stressed. A region becomes over-stressed either because more patients are coming in (more demand) or because there are fewer physicians available to treat them (less supply).

**Measuring stress absorb through patient flow network**. We measure how well a region (node) can **relieve healthcare stress** by redistributing excess demand to neighboring regions through the patient flow network. Given the excess stress $\lambda_i(t)$, after accounting for patient movement across regions, the remaining stress at region $i$ is

$$\lambda_i^W(t) = \lambda_i(t) - \sum_j w_{ij}(t) u_{ij}(t) \quad (4)$$

Where $w_{ij}(t)$ is the number of patients from region $i$ to region $j$. The term $u_{ij}(t) \in [0,1]$ represents the portion of stress that region $j$ is able to absorb, it defined as

$$u_{ij}(t) = \min\left(1, \frac{C_j(t)}{\sum_i w_{ij}(t)}\right) \quad (5)$$

The absorptivity for region $i$ is measured as

$$r_i(t) = 1 - \frac{\lambda_i^W(t)}{\lambda_i(t)} \quad (6)$$

A higher value of $r_i(t)$ indicate a higher portion of stress has been alleviated through patient flow network in the state. To measure overall system performance, we define state-wide absorptivity as:

$$r = 1 - \frac{\sum_i \sum_t \lambda_i^W(t)}{\sum_i \sum_t \lambda_i(t)} = 1 - \frac{\lambda^W(t)}{\lambda_i(t)} \quad (7)$$

**Measuring additional physician demand to achieve full absorptivity.** Although patient flows help redistribute demand, absorptivity is limited by available capacity $C_j(t)$). To achieve full absorptivity $r = 1$, we can increase the number of physicians in each region that expands the capacity for handling remaining stress after patient-flow network, that is

$$Q_i(t) = \frac{\lambda_i^W(t)}{\rho_i(t)} \quad (8)$$

where $Q_i(t)$ is the additional physician demand in region $i$ at time $t$. By comparing $Q_i(t)$ for different network configurations, we can know the reduction in physician demand achieved through optimized patient flow.

**Experiment settings.** We assess the absorptivity across regions under two distinct scenarios: (1) a real-world scenario, based on EHR data, and (2) a simulated broadcasting scenario, where regions proactively share available capacity with neighboring regions. The experimental setup is as follows:

• The real scenario: In this setting, we perform a comparative analysis of patient flow networks of the pre-pandemic period ($W^{Pre}$) and during the pandemic period ($W^{During}$). To ensure a fair comparison, we fix the same incoming patient volume $S_i(t)$ and regional treatment capacity $C_i(t)$ at each region $i$. By comparing the reduced stress between the two network configurations, we can identify whether during pandemic flow network presents superior performance in absorptivity.

• The simulated scenario: In this scenario, we assume that regions with available capacity can broadcast their ability to accept incoming patients to neighboring regions. The patient flow from region *i* to region *j* at time *t*, denoted $w_{ij}(t)$, is determined by the relative available treatment capacity of region *j* ( $C_j(t)$), defined as, $w_{ij}(t) = \frac{c_j(t)}{\sum_k \lambda_k(t)} \lambda_i(t)$. To simulate spatial constraints under 100km and 150 km, we restrict flows to occur only between regions within 100 km and 150 km distance thresholds. These simulations allow us to evaluate how constrained regional collaboration influences absorptivity under hypothetical network configurations.

**Acknowledgments**

The data, technology, and services used in generating these research findings were generously supplied pro bono by the COVID-19 Research Database partners, who are acknowledged at https://covid19researchdatabase.org/. We acknowledge the support of Research Accelerator grants funded by the Bill & Melinda Gates Foundation.

**Data availability**

The data that support the findings of this study were obtained under license from the COVID-19 Research Database consortium. The underlying EMR dataset is not publicly available. Our analyses were conducted exclusively on aggregated patient flow data, and at no point did we analyze individual-level data. All flows presented in this study were generated during the period in which we had valid access, prior to the permanent closure of the COVID-19 Research Database.

**Author contributions statement**

L.Z., S.P. and J.G. conceived the project and designed the study. L.Z. performed the data analyses. L.Z. and R.L. wrote the first draft of the manuscript. All authors contributed to interpreting the results and improving the manuscript.

**Competing interest statement**

The authors declare no competing interests.

## References

1. World Health Organization. *Operational Framework for Building Climate Resilient Health Systems*. 2015.
2. Sherman JD, MacNeill AJ, Biddinger PD, Ergun O, Salas RN, Eckelman MJ. Sustainable and resilient health care in the face of a changing climate. *Annu Rev Public Health*. 2023;44(1):255-277.
3. Haldane V, De Foo C, Abdalla SM, et al. Health systems resilience in managing the COVID-19 pandemic: lessons from 28 countries. *Nat Med*. 2021;27(6):964-980.
4. Zhong L, Lopez D, Pei S, Gao J. Healthcare system resilience and adaptability to pandemic disruptions in the United States. *Nat Med*. 2024;30(8):2311-2319.
5. Dyrbye LN, Shanafelt TD. Physician Burnout. *JAMA*. 2011;305(19). doi:10.1001/jama.2011.652
6. Krist AH, DeVoe JE, Cheng A, Ehrlich T, Jones SM. Redesigning Primary Care to Address the COVID-19 Pandemic in the Midst of the Pandemic. *The Annals of Family Medicine*. 2020;18(4):349-354. doi:10.1370/afm.2557
7. Miller IF, Becker AD, Grenfell BT, Metcalf CJE. Disease and healthcare burden of COVID-19 in the United States. *Nat Med*. 2020;26(8):1212-1217.
8. Pei S, Yamana TK, Kandula S, Galanti M, Shaman J. Burden and characteristics of COVID-19 in the United States during 2020. *Nature*. 2021;598(7880):338-341.
9. Overton CE, Pellis L, Stage HB, et al. EpiBeds: Data informed modelling of the COVID-19 hospital burden in England. *PLoS Comput Biol*. 2022;18(9):e1010406.
10. Villarroel L, Christ CM, Smith L, et al. Collaboration on the Arizona surge line: how Covid-19 became the impetus for public, private, and federal hospitals to function as one system. *NEJM Catal Innov Care Deliv*. 2021;2(1).
11. Schaye VE, Reich JA, Bosworth BP, et al. Collaborating across private, public, community, and federal hospital systems: lessons learned from the Covid-19 pandemic response in NYC. *NEJM Catal Innov Care Deliv*. 2020;1(6).
12. Pett E, Leung HL, Taylor E, et al. Critical care transfers and COVID-19: managing capacity challenges through critical care networks. *J Intensive Care Soc*. 2022;23(2):203-209.
13. Mitchell SH, Rigler J, Baum K. Regional transfer coordination and hospital load balancing during COVID-19 surges. In: *JAMA Health Forum*. Vol 3. 2022:e215048–e215048.
14. Della Rossa F, Salzano D, Di Meglio A, et al. A network model of Italy shows that intermittent regional strategies can alleviate the COVID-19 epidemic. *Nat Commun*. 2020;11(1):5106. doi:10.1038/s41467-020-18827-5
15. Sakr CJ, Assaf SA, Fakih L, et al. Hospitals' Collaborations Strengthen Pandemic Preparedness: Lessons Learnt from COVID-19. In: *Healthcare*. Vol 12. 2024:321.
16. Hick JL, Hanfling D, Burstein JL, et al. Health care facility and community strategies for patient care surge capacity. *Ann Emerg Med*. 2004;44(3):253-261. doi:10.1016/j.annemergmed.2004.04.011
17. Bergmark RW, Jin G, Semco RS, Santolini M, Olsen MA, Dhand A. Association of hospital centrality in inter-hospital patient-sharing networks with patient mortality and length of stay. *PLoS One*. 2023;18(3):e0281871. doi:10.1371/journal.pone.0281871
18. Ceferino L, Mitrani-Reiser J, Kiremidjian A, Deierlein G, Bambarén C. Effective plans for hospital system response to earthquake emergencies. *Nat Commun*. 2020;11(1):4325.
19. Agostini L, Onofrio R, Piccolo C, Stefanini A. A management perspective on resilience in healthcare: a framework and avenues for future research. *BMC Health Serv Res*. 2023;23(1):774. doi:10.1186/s12913-023-09701-3
20. Arsenault C, Gage A, Kim MK, et al. COVID-19 and resilience of healthcare systems in ten countries. *Nat Med*. 2022;28(6):1314-1324.
21. Kruk ME, Ling EJ, Bitton A, et al. Building resilient health systems: a proposal for a resilience index. *BMJ*. Published online May 23, 2017:j2323. doi:10.1136/bmj.j2323
22. West CP, Dyrbye LN, Sinsky C, et al. Resilience and Burnout Among Physicians and the General US Working Population. *JAMA Netw Open*. 2020;3(7):e209385. doi:10.1001/jamanetworkopen.2020.9385
23. Burau V, Falkenbach M, Neri S, Peckham S, Wallenburg I, Kuhlmann E. Health system resilience and health workforce capacities: Comparing health system responses during the COVID-19 pandemic in six European countries. *Int J Health Plann Manage*. 2022;37(4):2032-2048. doi:10.1002/hpm.3446

24. Austin JM, Kachalia A. The state of health care quality measurement in the era of COVID-19: the importance of doing better. *JAMA*. 2020;324(4):333-334.
25. Akachi Y, Kruk ME. Quality of care: measuring a neglected driver of improved health. *Bull World Health Organ*. 2017;95(6):465-472. doi:10.2471/BLT.16.180190
26. Zachrison KS, Onnela J, Reeves MJ, et al. Hospital Factors Associated With Interhospital Transfer Destination for Stroke in the Northeast United States. *J Am Heart Assoc*. 2020;9(1). doi:10.1161/JAHA.118.011575
27. Tong L, Patel R V., Aizer AA, Dhand A, Bi WL. Role of Hospital Connectedness in Brain Metastasis Outcomes. *JAMA Netw Open*. 2024;7(9):e2435051. doi:10.1001/jamanetworkopen.2024.35051
28. Liu X, Li D, Ma M, Szymanski BK, Stanley HE, Gao J. Network resilience. *Phys Rep*. 2022;971:1-108.
29. Gao J, Barzel B, Barabási AL. Universal resilience patterns in complex networks. *Nature*. 2016;530(7590):307-312.
30. Hassan EM, Mahmoud HN. Orchestrating performance of healthcare networks subjected to the compound events of natural disasters and pandemic. *Nat Commun*. 2021;12(1):1338.
31. Lo Sardo DR, Thurner S, Sorger J, Duftschmid G, Endel G, Klimek P. Quantification of the resilience of primary care networks by stress testing the health care system. *Proceedings of the National Academy of Sciences*. 2019;116(48):23930-23935.
32. Tien JM, Goldschmidt-Clermont PJ. Healthcare: A complex service system. *J Syst Sci Syst Eng*. 2009;18(3):257-282.
33. Zhang H, Liu X, Wang Q, Zhang W, Gao J. Co-adaptation enhances the resilience of mutualistic networks. *J R Soc Interface*. 2020;17(168):20200236.
34. Linkov I, Trump BD. *The Science and Practice of Resilience*. Springer; 2019.
35. Yabe T, Rao PSC, Ukkusuri S V, Cutter SL. Toward data-driven, dynamical complex systems approaches to disaster resilience. *Proceedings of the National Academy of Sciences*. 2022;119(8):e2111997119.
36. Haldane V, Morgan GT. From resilient to transilient health systems: the deep transformation of health systems in response to the COVID-19 pandemic. *Health Policy Plan*. 2021;36(1):134-135. doi:10.1093/heapol/czaa169
37. Saulnier DD, Blanchet K, Canila C, et al. A health systems resilience research agenda: moving from concept to practice. *BMJ Glob Health*. 2021;6(8):e006779.
38. Barnett ML, Christakis NA, O'Malley J, Onnela JP, Keating NL, Landon BE. Physician Patient-sharing Networks and the Cost and Intensity of Care in US Hospitals. *Med Care*. 2012;50(2):152-160. doi:10.1097/MLR.0b013e31822dcef7
39. Alexander GC, Tajanlangit M, Heyward J, Mansour O, Qato DM, Stafford RS. Use and Content of Primary Care Office-Based vs Telemedicine Care Visits During the COVID-19 Pandemic in the US. *JAMA Netw Open*. 2020;3(10):e2021476. doi:10.1001/jamanetworkopen.2020.21476
40. Bourgeault IL, Maier CB, Dieleman M, et al. The COVID-19 pandemic presents an opportunity to develop more sustainable health workforces. *Hum Resour Health*. 2020;18(1):83. doi:10.1186/s12960-020-00529-0
41. O'Malley AS, Reschovsky JD. Referral and Consultation Communication Between Primary Care and Specialist Physicians. *Arch Intern Med*. 2011;171(1). doi:10.1001/archinternmed.2010.480
42. Dahlen A, Charu V. Analysis of sampling bias in large health care claims databases. *JAMA Netw Open*. 2023;6(1):e2249804–e2249804.
43. Seung-Chul K, Ira H, others. Flexible bed allocation and performance in the intensive care unit. *Journal of Operations Management*. 2000;18(4):427-443.
44. Taylor A, Murakami M, Kim S, Chu R, Riek LD. Hospitals of the Future: Designing Interactive Robotic Systems for Resilient Emergency Departments. *Proc ACM Hum Comput Interact*. 2022;6(CSCW2):1-40. doi:10.1145/3555543
45. The COVID-19 Research Database . Accessed February 19, 2026. https://hhstechgroup.com/covid-19-research-database-partners-with-the-hhs-technology-group-scale-an-open-research-database/

**Table 1: Characteristics associated with cross-regional flows.** Values represent the mean across 460 three-digit ZIP code regions, with ranges indicating 95% confidence intervals. Differences between the pre-pandemic (2017–2019) and during-pandemic (2020–2022) networks were evaluated using two-sided t-tests and considered statistically significant at $P < 0.05$.

| Characteristics | Pre-pandemic 2017-2019 | During-pandemic 2020-2022 | Difference |
|---|---|---|---|
| | Mean [Range] | Mean [Range] | Mean Diff [P-value] |
| **Cross-regional flow** $\sigma$ (Percent) | 2.08 [1.62-2.54] | 2.81 [2.34-3.29] | **0.74 [P=0.0021]** |
| Small Region | 5.99 [0.20-11.79] | 5.58 [0.49-10.67] | -0.41[P=0.6656] |
| Medium Region | 5.59 [1.77-9.40] | 3.74 [2.00-5.48] | -1.84[P=0.8135] |
| Large Region | 1.68 [1.25-2.10] | 2.56 [2.04-3.07] | **0.88[P=0.0352]** |
| Mega Region | 1.89 [0.86-2.92] | 3.40 [1.92-4.89] | **1.52[P=0.0066]** |
| **Out-degree** | 1.05 [0.93-1.17] | 1.13 [1.00-1.25] | **0.08 [P=0.0027]** |
| Small Region | 2.00 [0.66-3.34] | 1.60 [0.67-2.54] | -0.40[P=0.6041] |
| Medium Region | 1.40 [0.71-2.09] | 1.05 [0.64-1.45] | -0.35[P=0.1324] |
| Large Region | 0.99 [0.86-1.11] | 1.02 [0.90-1.15] | 0.03[P=0.0620] |
| Mega Region | 1.06 [0.73-1.38] | 1.67 [1.18-2.15] | **0.61[P=0.0014]** |
| **Distance** | 38.34 [33.18-43.50] | 42.07 [37.86-46.28] | **3.73 [P=0.0000]** |
| Small Region | 53.33 [22.68-83.98] | 56.39 [23.73-89.04] | 3.06[P=0.5533] |
| Medium Region | 62.65 [28.62-96.69] | 54.20 [29.99-78.41] | -8.45[P=0.7533] |
| Large Region | 37.70 [31.91-43.48] | 41.44 [36.62-46.26] | **3.75[P=0.0002]** |
| Mega Region | 28.12 [19.39-36.86] | 37.81 [29.95-45.68] | **9.69[P=0.0014]** |
| | Pre-pandemic Network | During-pandemic Network | |
| | Mean [Range] | Mean [Range] | Mean Diff [P-value] |
| Absorptivity | 0.15 [0.14-0.16] | 0.58 [0.57-0.60] | **0.43 [P=0.0000]** |
| Small Region | 0.26 [0.20-0.32] | 0.56 [0.47-0.66] | **0.31[P=0.0000]** |
| Medium Region | 0.21 [0.17-0.24] | 0.60 [0.54-0.67] | **0.40[P=0.0000]** |
| Large Region | 0.15 [0.14-0.16] | 0.57 [0.55-0.59] | **0.42[P=0.0000]** |
| Mega Region | 0.12 [0.10-0.14] | 0.66 [0.62-0.70] | **0.54[P=0.0000]** |

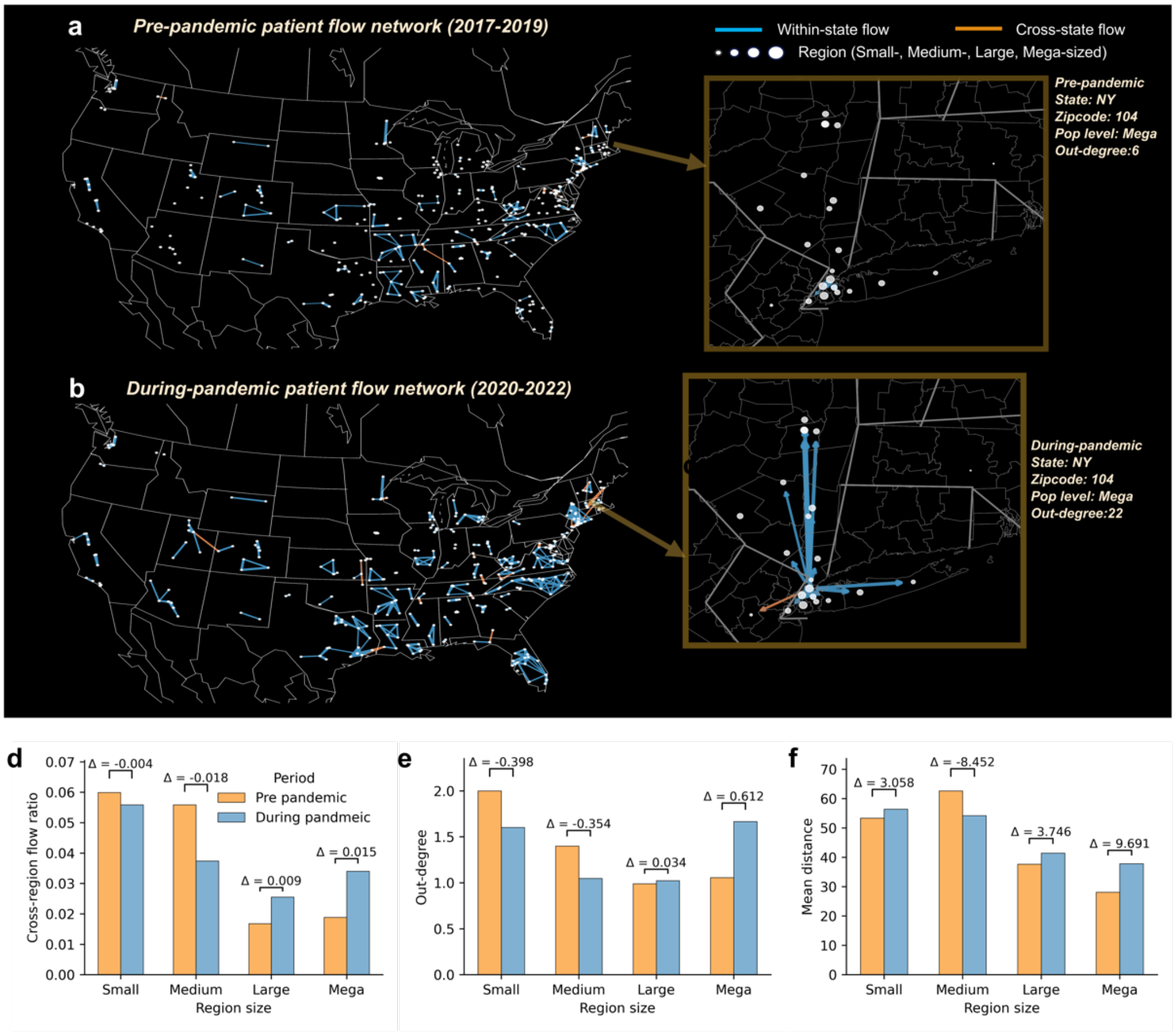

**Figure 1. Structural shifts in the U.S. patient flow network before and during the COVID-19 pandemic.** Patient movements between sub-regions (three-digit ZIP codes) were used to construct inter-regional flow networks for two periods: pre-pandemic (2017–2019) and during-pandemic (2020–2022). **(a,b)** Spatial distribution of inter-regional patient flow networks in the pre-pandemic and during-pandemic periods. Nodes represent sub-regions, and edges represent patient flows, with thicker edges indicating larger flow volumes. Insets show examples of outgoing flows from selected sub-regions in New York. Regions are categorized by population size as small, medium, large, and mega. **(c)** Proportion of cross-regional flow by region size. **(d)** Mean out-degree (number of outgoing connections) by region size. **(e)** Mean travel distance between connected sub-regions by region size. Compared with the pre-pandemic period, the during-pandemic

network exhibits higher cross-regional flow, increased connectivity, and longer travel distances, indicating expanded inter-regional patient movements across states.

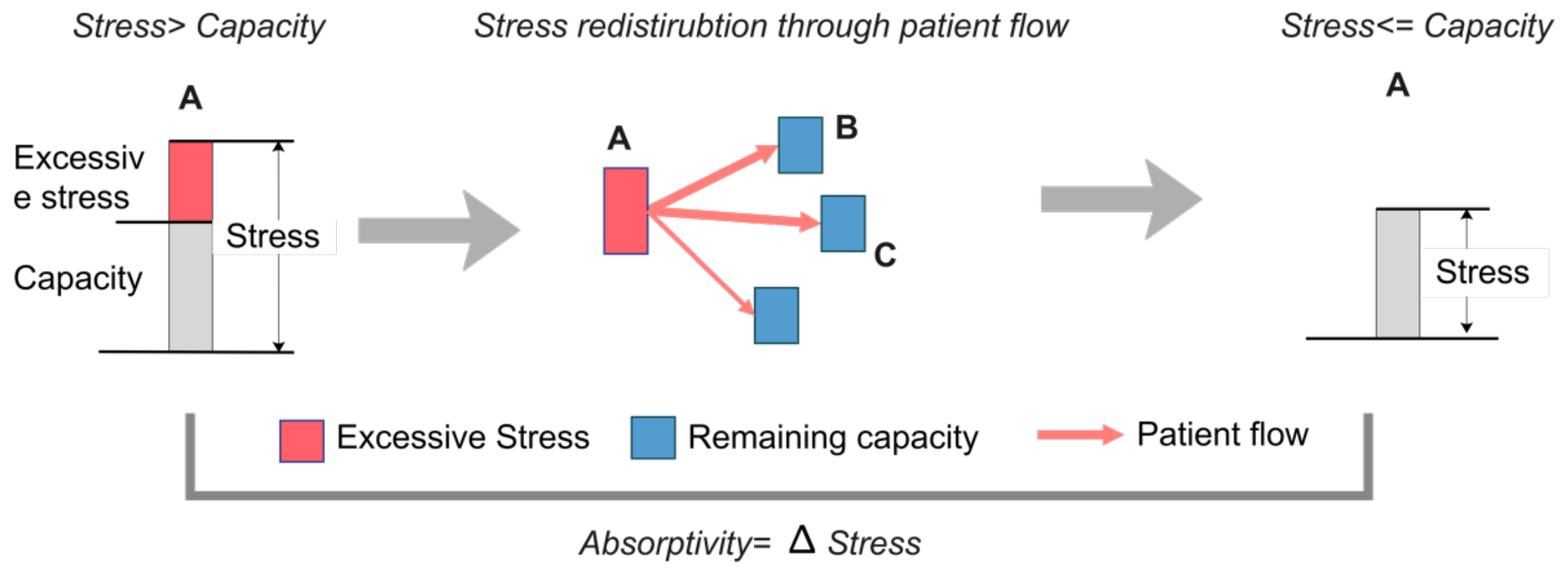

**Figure 2: Absorptivity as a measure of structural resilience.** For a focal region (node A), when stress exceeds its local capacity, the excessive stress is redistributed to neighboring regions (nodes B, C, and D) through patient flows/ This redistribution reduces the stress burden on the originating region as neighboring regions absorb part of the excess demand. Absorptivity is defined as the magnitude of stress reduction achieved through these patient flows, reflecting the network's capacity to mitigate localized overload via inter-regional redistribution.

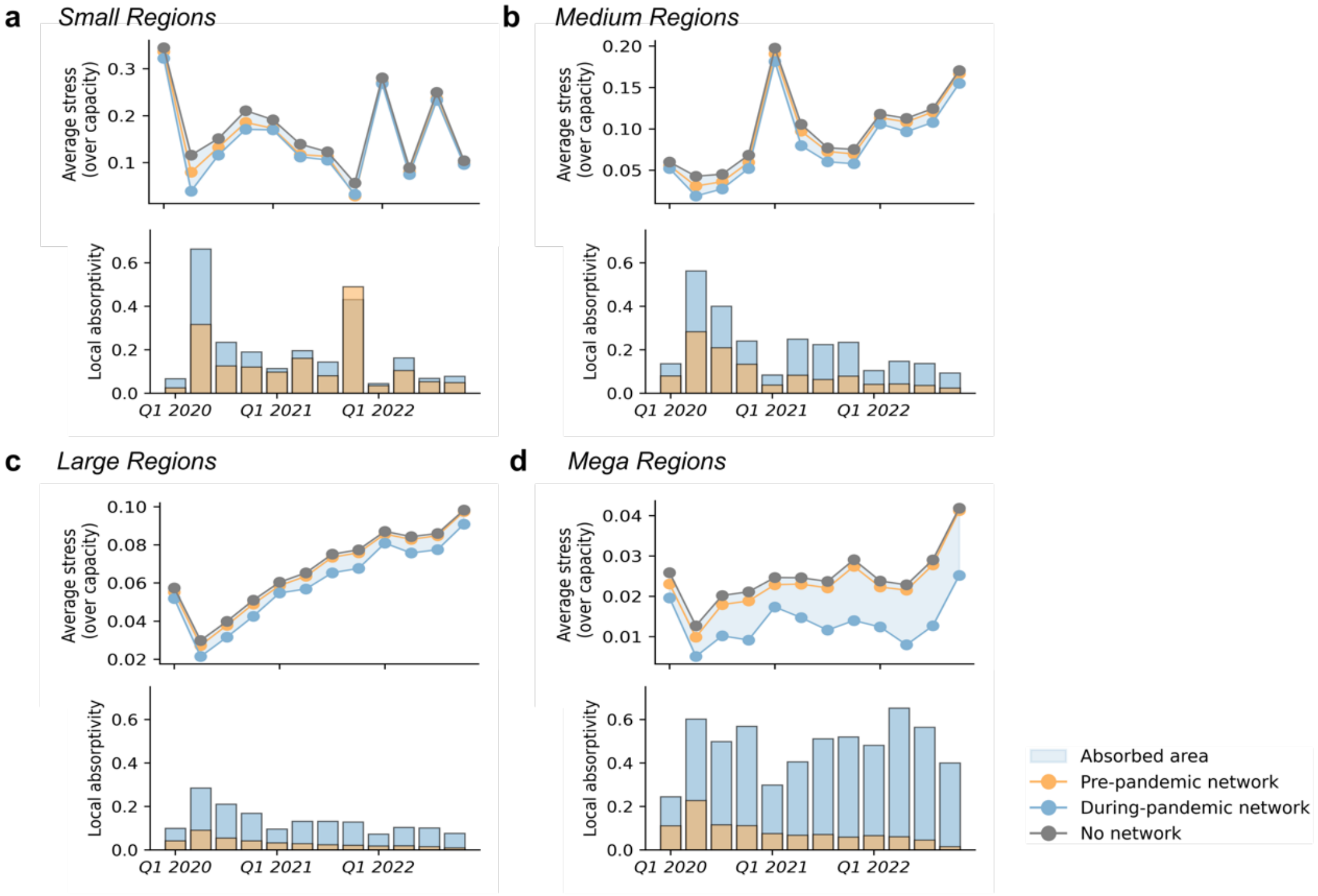

**Figure 3. Temporal patterns of excess stress and absorptivity across region sizes.** Seasonally excess healthcare stress, normalized by local treatment capacity, for (**a**) small-, (**b**) medium-, (**c**) large-, and (**d**) mega-sized regions. Lower panels show corresponding seasonal absorptivity across regions. By constructing a counterfactual scenario using the pre-pandemic (2017–2019) patient flow network as a fixed baseline we show that during-pandemic networks consistently exhibit higher absorptivity, indicating an enhanced capacity to redistribute excess healthcare demand.

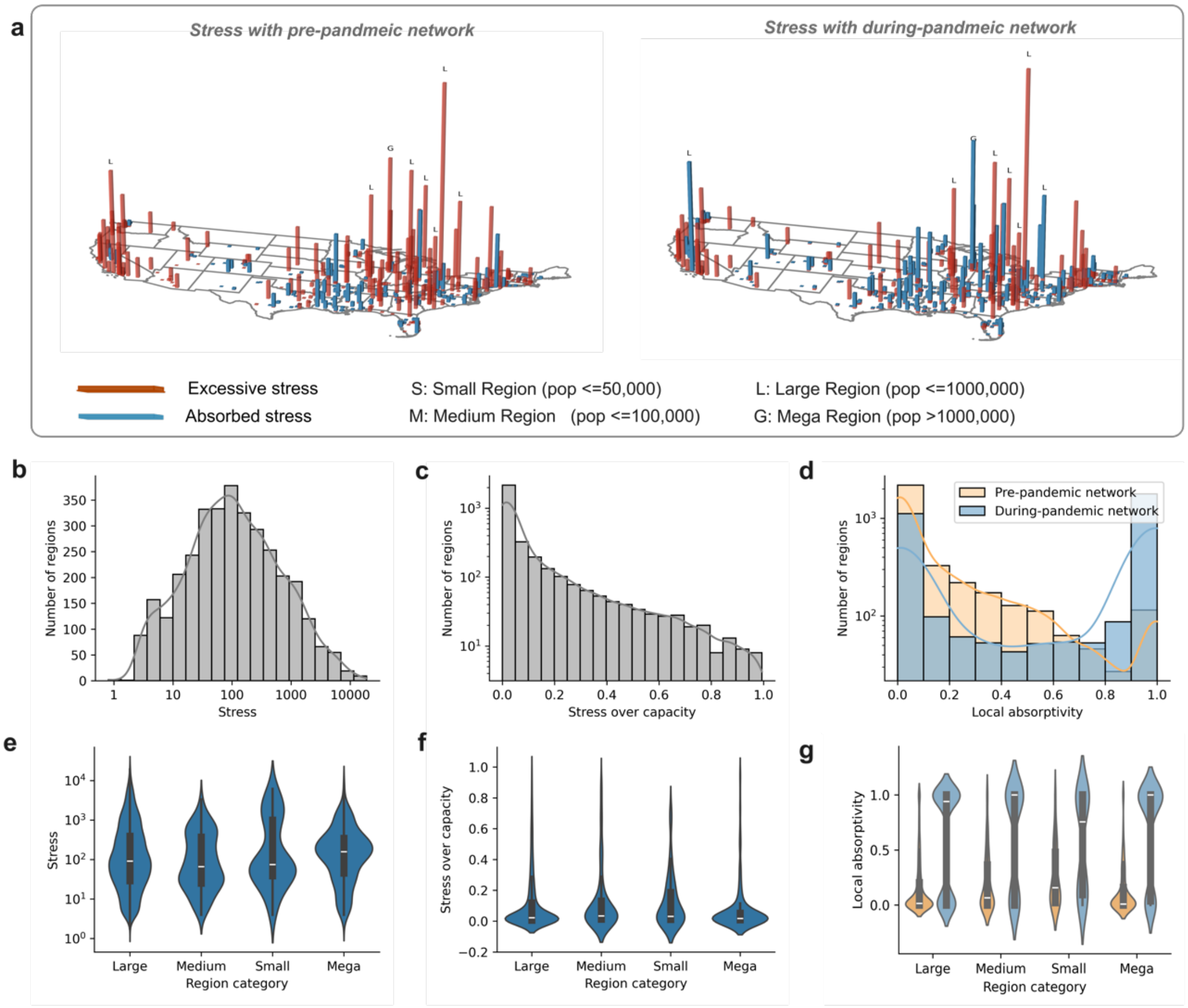

**Figure 4: Spatial distribution of excessive stress and absorptivity**. **(a)** Spatial distribution of stress across sub-regions. **(b)** Distribution of raw stress levels across all sub-regions. **(c)** Distribution of excess stress normalized by local treatment capacity. **(d)** Distribution of absorptivity across sub-regions, comparing pre-pandemic and during-pandemic patient flow networks . **(e–g)** Stratified analyses by region population size (Small, Medium, Large, and Mega): **(e)** distribution of raw excess stress; **(f)** excess stress relative to treatment capacity; and **(g)** distribution of absorptivity under pre-pandemic and during-pandemic network configurations.

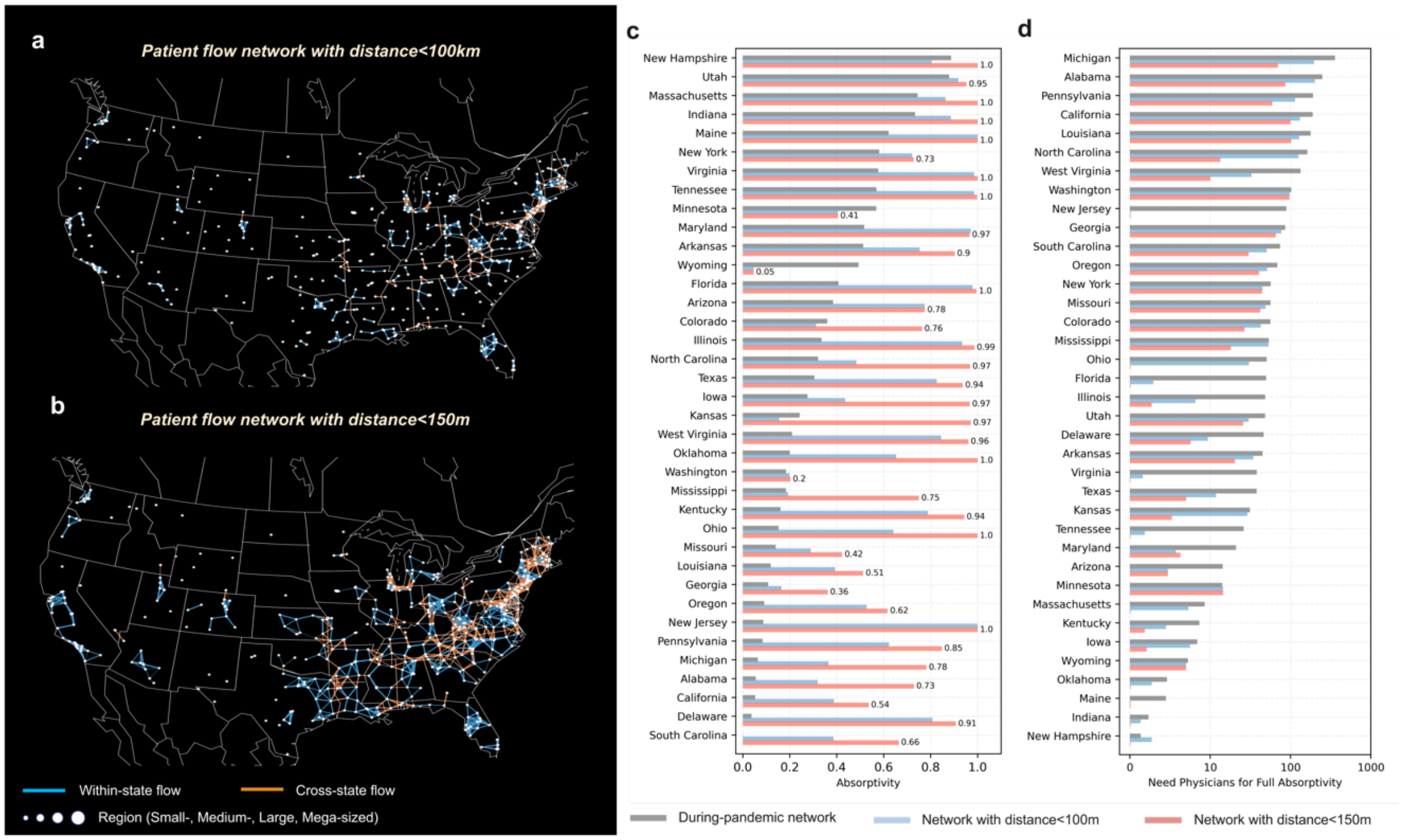

**Figure 5: Regional collaboration enhances absorptivity and reduces physician demand.** Assuming regional collaboration, regions with available capacity broadcast to neighboring regions under stress, subject to maximum distance thresholds of 100 km and 150 km. **(a, b)** Spatial distribution of the flow networks under the two thresholds. **(c)** State-level absorptivity under the collaborative networks of two thresholds. **(d)** Additional number of physicians required in each state to reach full absorptivity (r = 1) under the collaborative networks.

**Supplementary Information**

**Dataset.** Table S1-S2 presents essential statistics for the Healthjump datasets spanning from 2017 to 2022, including the total number of records and the proportion of valid records among various age and race groups. Valid data points are notably high at 99.8% across age groups, moderate at 58.2% across race groups.

To ensure each state is adequately represented in the dataset, we exclude those with insufficient data, defined as states where patient numbers are less than 0.1% of the state population and include fewer than three subregions. The excluded states are AK, CT, HI, ID, MT, NE, NV, NM, ND, RI, SD, VT and WI. Table S2 presents the statistics for the included states, details the characteristics of their cross-regional flow networks, and the results of network absorptivity.

**Imbalances in Healthcare Demand and Physician Capacity.** Figure S1 a-c presents trends in patient volume, physician counts, and the patient-to-physician ratio across U.S.\ states from 2016 to 2022 in our dataset. Consistent with prior studies, total patient volume dropped markedly during the pandemic. However, the number of primary care physicians (PCPs) also declined relative to expected trends. Under normal circumstances, physician counts in the dataset would have continued to increase due to expanded electronic health record (EHR) coverage. Instead, workforce participation decreased. As shown in Fig. S1 c, the patient-to-physician ratio fell in 2020, reflecting the initial drop in utilization, but then increased throughout 2021–2022. This rebound occurred because physician supply did not recover at the same pace as patient demand. In other words, although fewer patients sought care overall, the reduction in the primary care workforce resulted in increased workload per remaining physician in the post-2020 period. This pattern is consistent with the shortage of the physician workforce in the 2024 Association of American Medical Colleges report.

Figure S2 illustrates the spatial and temporal distribution of excessive stress from 2020 to 2022. Although the national level appears lower in early 2020 in Fig. S1c, a more granular regional analysis in Figure S2 reveals that some regions experienced substantial surges in excessive stress even during periods of relatively low overall healthcare utilization.

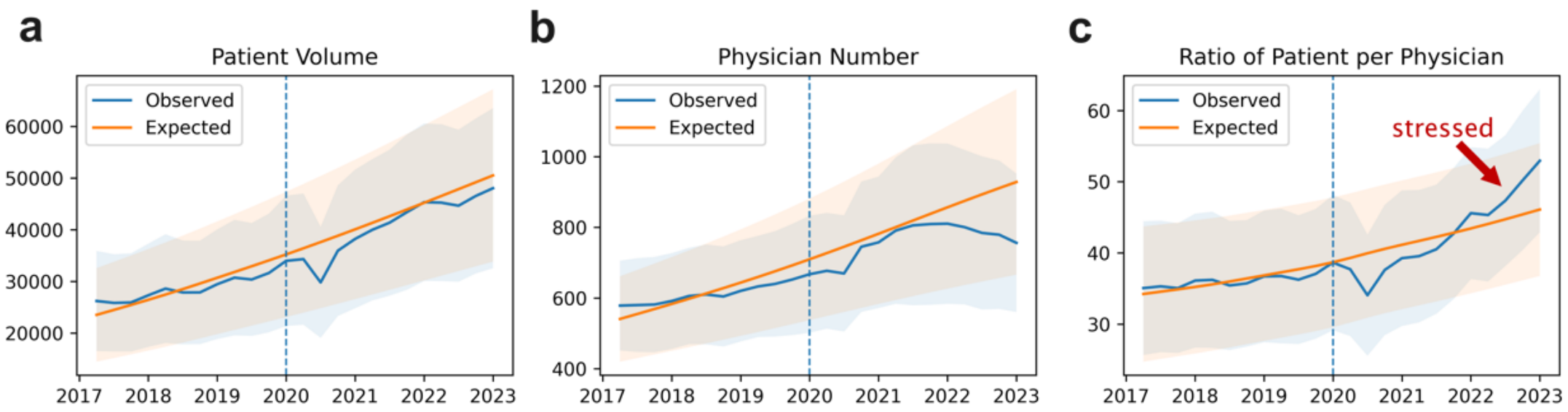

**Figure S1. Trends in patient volume, physician counts, and patient-to-physician ratios across U.S. states in primary care.** (**a**) Trend of patient volume. (**b**) Trend in the number of primary care physicians. (**c**) Trends in the patient-to-physician ratio. Expected patient volume and physician counts are estimated using logistic regression models fitted to pre-pandemic data.

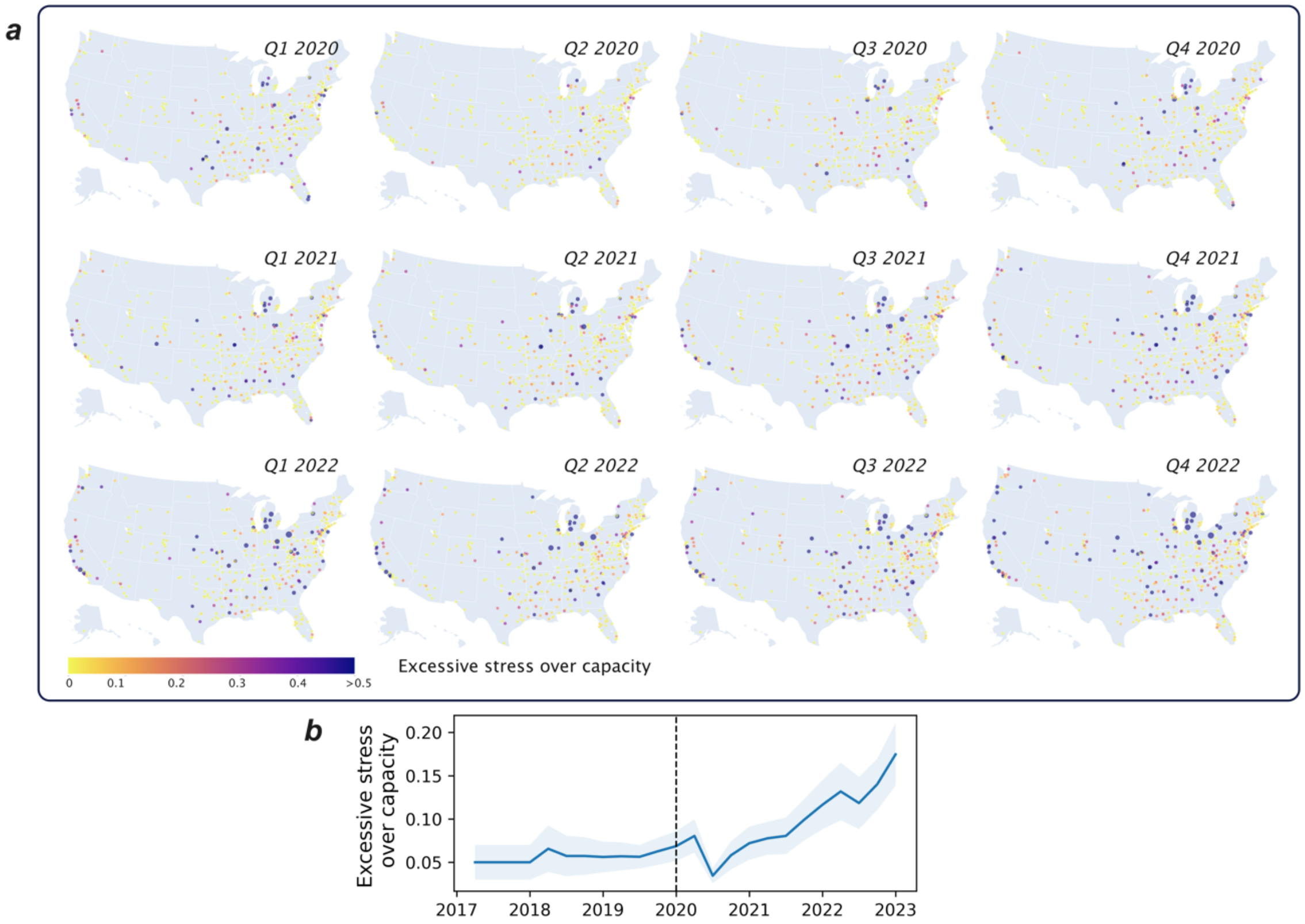

**Figure S2. Excessive healthcare stress relative to capacity across U.S. regions.** (**a**) Spatial distribution of excessive stress, defined as the extent to which observed patient demand exceeds estimated primary care capacity. Each circle represents a three-digit region, with larger sizes and darker colors indicating higher levels of excessive stress. (**b**) Temporal patterns of excessive stress across regions.

Table S1: Total number of electronic health records for primary care in the dataset.

| Number of records | 2017 | 2018 | 2019 | 2020 | 2021 | 2022 | Total |
|---|---|---|---|---|---|---|---|
| Primary care | 8,813,951 | 9,507,456 | 10,528,706 | 11,415,625 | 14,087,571 | 15,636,557 | 69,989,866 |

Table S2: Statistics of datasets across states and their patient networks. States (including AK, AZ, CT, HI, ID, ME, MI, MT, NE, NV, NM, ND, OK, RI, SD, VT, and WI) were not analyzed due to insufficient data, with the number of sub-regions is less than 3.

| State | Period | Number of pa-patients (Percentage of total population) | Number of sub-regions (Total sub-regions) | Percentage of cross-region flows | Topological measures | | | Absorptivity |
|---|---|---|---|---|---|---|---|---|
| | | | | | Average distance (km) | Density | Heterogeneity | COVID-19 pandemic |
| AL | Pre-pandemic | 160, 412 (3.3%) | 12(19) | 1.05 | 60 | 0.03 | 0.17 | 0.02 |
| | During-pandemic | 196,732 (4.0%) | 12(19) | 0.93 | 57 | 0.03 | 0.14 | 0.04 |
| AR | Pre-pandemic | 76,665 (2.6%) | 12(14) | 0.62 | 111 | 0.1 | 0.31 | 0.03 |
| | During-pandemic | 129,512 (4.3%) | 12(14) | 1.95 | 104 | 0.13 | 0.83 | 0.31 |
| CA | Pre-pandemic | 149,148 (0.4%) | 21(57) | 0.66 | 81 | 0.004 | 0 | 0.01 |
| | During-pandemic | 209,001 (0.5%) | 23 (57) | 0.93 | 78 | 0.004 | 0.09 | 0.03 |
| CO | Pre-pandemic | 91,784 (1.7%) | 13(17) | 0.29 | 121 | 0.04 | 0.09 | 0.02 |
| | During-pandemic | 131,759 (2.4%) | 12 (17) | 1.06 | 132 | 0.04 | 0.13 | 0.16 |
| DE | Pre-pandemic | 37,085 (3.9 %) | 3(3) | 0.34 | 14 | 0.33 | 0 | 0.01 |
| | During-pandemic | 38355 (4.0 %) | 3(3) | 0.21 | 14 | 0.33 | 0 | 0.01 |
| FL | Pre-pandemic | 123,982 (0.6 %) | 18 (25) | 1.07 | 47 | 0.02 | 0.57 | 0.08 |
| | During-pandemic | 231,274 (1.1%) | 20 (25) | 1.47 | 64 | 0.05 | 0.79 | 0.4 |
| GA | Pre-pandemic | 61463 (0.6%) | 12(20) | 1.33 | 47 | 0.03 | 0.62 | 0.03 |
| | During-pandemic | 81610 (0.8%) | 11(20) | 1.5 | 47 | 0.02 | 0.51 | 0.11 |
| IL | Pre-pandemic | 38,278 (0.3%) | 9(29) | 1.27 | 34 | 0 | 0.25 | 0.02 |
| | During-pandemic | 81,949 (0.6%) | 11(29) | 2.26 | 77 | 0.02 | 0.4 | 0.24 |
| IN | Pre-pandemic | 10,428 (0.2 %) | 6(20) | 0.24 | 31 | 0.01 | 0 | 0.03 |
| | During-pandemic | 42,127 (0.6 %) | 6 (20) | 0.43 | 34 | 0.01 | 0 | 0.65 |
| IA | Pre-pandemic | 7,185 (0.2%) | 7(25) | 0.38 | 15 | 0.02 | 0.07 | 0.02 |
| | During-pandemic | 8,734 (0.3%) | 7(25) | 0.56 | 18 | 0.01 | 0.21 | 0.09 |
| KS | Pre-pandemic | 23,140 (0.8%) | 9(19) | 0.24 | 95 | 0.03 | 0.41 | 0.01 |
| | During-pandemic | 19,685 (0.7%) | 10 (19) | 0.39 | 120 | 0.02 | 0.46 | 0.04 |
| KY | Pre-pandemic | 5,490 (0.1%) | 7(27) | 0.04 | 25 | 0 | 0.5 | 0 |
| | During-pandemic | 14,010 (0.3%) | 7(27) | 0.73 | 85 | 0 | 0.48 | 0.04 |
| LA | Pre-pandemic | 96946 (2.1%) | 13 (13) | 0.63 | 93 | 0.19 | 0.49 | 0.01 |
| | During-pandemic | 136009 (2.9%) | 13 (13) | 1.18 | 95 | 0.17 | 0.48 | 0.08 |
| MD | Pre-pandemic | 41616 (7%) | 10 (13) | 0.77 | 32 | 0.04 | 0.17 | 0.07 |
| | During-pandemic | 47821 (8%) | 8 (13) | 0.49 | 30 | 0.03 | 0.24 | 0.15 |
| MA | Pre-pandemic | 12368 (0.2%) | 3 (18) | 4.94 | 36 | 0.01 | 0 | 0.14 |
| | During-pandemic | 17728 (0.3%) | 6 (18) | 7.31 | 92 | 0.02 | 0.28 | 0.72 |
| MN | Pre-pandemic | 8106 (0.1%) | 3 (16) | 3.51 | 238 | 0.01 | 0 | 0.08 |
| | During-pandemic | 11280 (0.2%) | 3 (16) | 2.98 | 164 | 0.01 | 0.17 | 0.21 |
| MS | Pre-pandemic | 93393 (3.1%) | 8 (12) | 1.18 | 113 | 0.08 | 0.37 | 0.04 |
| | During-pandemic | 96002 (3.2%) | 8 (12) | 1.94 | 112 | 0.07 | 0.28 | 0.13 |
| MO | Pre-pandemic | 22487 (0.4%) | 7 (25) | 2.16 | 79 | 0.02 | 0.33 | 0.02 |
| | During-pandemic | 26485 (0.4%) | 7 (25) | 2.29 | 74 | 0.02 | 0.25 | 0.06 |
| NH | Pre-pandemic | 7395 (0.6%) | 3 (9) | 4.73 | 75 | 0.08 | 0 | 0.48 |
| | During-pandemic | 9574 (0.7%) | 3 (9) | 3.61 | 78 | 0.08 | 0 | 0.88 |
| NJ | Pre-pandemic | 88332 (1%) | 4 (20) | 1.22 | 34 | 0.01 | 0.13 | 0.02 |
| | During-pandemic | 108886 (1.2%) | 4 (20) | 1.67 | 21 | 0.01 | 0 | 0.08 |
| NY | Pre-pandemic | 26104 (0.1%) | 17 (51) | 3.54 | 44 | 0.01 | 0.67 | 0.05 |
| | During-pandemic | 57029 (0.3%) | 22 (51) | 6.7 | 82 | 0.03 | 1.54 | 0.64 |
| NC | Pre-pandemic | 255189 (2.5%) | 20 (20) | 0.37 | 75 | 0.08 | 0.9 | 0.02 |
| | During-pandemic | 373660 (3.7%) | 20 (20) | 0.74 | 87 | 0.09 | 1.05 | 0.14 |
| OH | Pre-pandemic | 42039 (0.4%) | 10 (29) | 2.1 | 48 | 0.01 | 0.32 | 0.05 |
| | During-pandemic | 57721 (0.5%) | 11 (20) | 1.58 | 42 | 0.01 | 0.22 | 0.13 |
| PA | Pre-pandemic | 91912 (0.7%) | 11 (46) | 0.01 | 44 | 0.001 | 0 | 0.001 |
| | During-pandemic | 106391 (0.8%) | 10 (46) | 0.06 | 47 | 0.001 | 0.07 | 0.004 |
| SC | Pre-pandemic | 22654 (0.5%) | 4 (10) | 0 | 64 | 0.01 | 0.5 | 0 |
| | During-pandemic | 31313 (0.6%) | 5 (10) | 0.06 | 64 | 0.02 | 0.31 | 0.006 |
| TN | Pre-pandemic | 66789 (1%) | 11 (15) | 1.45 | 29 | 0.05 | 0.09 | 0.06 |
| | During-pandemic | 129413 (1.9%) | 12 (15) | 2.08 | 42 | 0.07 | 0.41 | 0.49 |
| TX | Pre-pandemic | 100139 (0.4%) | 24 (50) | 0.7 | 54 | 0.01 | 0.17 | 0.03 |
| | During-pandemic | 205738 (0.7%) | 26 (50) | 1.38 | 72 | 0.01 | 0.44 | 0.3 |
| UT | Pre-pandemic | 286013 (9,4%) | 5 (7) | 1.52 | 62 | 0.1 | 0.17 | 0.23 |
| | During-pandemic | 310926 (10.2%) | 5 (7) | 1.51 | 91 | 0.13 | 0.24 | 0.56 |
| VA | Pre-pandemic | 81001 (1%) | 21 (28) | 0.98 | 44 | 0.04 | 0.41 | 0.07 |
| | During-pandemic | 124147 (1.5%) | 24 (28) | 1.71 | 58 | 0.05 | 0.54 | 0.44 |
| WA | Pre-pandemic | 77551 (1.1%) | 5 (14) | 0.63 | 82 | 0.01 | 0.33 | 0.02 |
| | During-pandemic | 94825 (1.3%) | 4 (14) | 0.78 | 82 | 0.02 | 0.11 | 0.06 |
| WV | Pre-pandemic | 67672 (3.7%) | 14 (22) | 8.18 | 56 | 0.06 | 0.94 | 0.09 |
| | During-pandemic | 80641 (4.4%) | 14 (22) | 8.09 | 56 | 0.04 | 0.73 | 0.16 |
| WY | Pre-pandemic | 1975 (0.3%) | 3 (13) | 0.63 | 280 | 0.01 | 0 | 0.004 |
| | During-pandemic | 3570 (0.6%) | 4 (13) | 0.69 | 280 | 0.01 | 0 | 0.06 |